\documentstyle[preprint,aps]{revtex}

\begin{document}
\draft
\title{Investigation of Long Baseline Neutrino Oscillations from
Experiments at KAON}

\author{Scott Hayward}
\address{Department of Physics, University of British Columbia,
Vancouver, B.C. V6T 2A6, Canada }
\author{Martin Sevior}
\address{School of Physics, The University of Melbourne,
Parkville, Vic., 3052, Australia }
\author{Nathan Weiss}
\address{Department of Physics, University of British Columbia,
Vancouver, B.C. V6T 2A6, Canada }
\author{Dennis Wright}
\address{Triumf, 4004 Wesbrook Mall, Vancouver, B.C. V6T 2A3, Canada}

\maketitle

\begin{abstract}
The proposed KAON factory at the Triumf Laboratory in Vancouver, Canada
provides
a unique opportunity for high statistics long baseline neutrino oscillation
experiments. Several possibilities are under active consideration. In this
paper
we describe the theoretical expectations for a {\it very} long baseline
experiment in which the neutrinos are directed towards, and detected at the
Superkamiokande detector in Kamioka, Japan 7200 km away. We find that in the
first year this experiment would probe oscillations down to about $\Delta m^2_0
= 9 \times  10^{-5} eV^2$ for maximal mixing, and if $\Delta m^2_0 \geq 5
\times
10^{-4} eV^2$ it would be sensitive to $sin^2(2 \theta_0) \geq 0.2$. These
results are compared with a more modest proposal of a 100 km baseline with a
6300 tonne detector which would probe $\Delta m^2_0 \geq 1.3 \times 10^{-3}
eV^2$ for maximal mixing, and if $\Delta m^2_0 \geq 0.02 eV^2$ it would be
sensitive to $sin^2(2 \theta_0) \geq 0.01$. These experiments would either
confirm or rule out the entire range of parameters allowed by  Kamiokande and
IMB  to explain the deficit in the ratio of $\nu_\mu$ to $\nu_e$ in current
atmospheric neutrino experiments. The Kamiokande experiment would also
investigate matter enhanced neutrino oscillations (the MSW effect) down to
about
$\Delta m^2 = 3 \times 10^{-4} eV^2$.
\end{abstract}

\pacs{ }

\narrowtext
\section{Introduction}
\label{sec:intro}
In the past decade neutrino physics has become a very active field partly
as a result of the Solar Neutrino problem. Several new detectors including
Kamiokande, SAGE, and Gallex have confirmed earlier results from the Homestake
experiment that there is a deficit in the number of solar neutrinos arriving
on the earth. More recently Kamiokande and IMB have reported deficits in
the expected ratio of muon to electron type neutrinos produced in the
decays of pions and kaons in cosmic ray
showers which occur in the upper atmosphere. This latter
effect has been referred to as the Atmospheric Neutrino problem.

The deficit in the number of solar neutrinos may either require modification
of the
standard solar model or it may be due to new physics beyond the standard model
of particle physics. It is clear \cite{SSM} that minor changes to the standard
solar model cannot account for the large deficits which are seen.
Furthermore, as has been pointed out by Bachall \cite{JNB1,JNB-HAB}, there is
direct
experimental evidence that the neutrino spectrum is affected by the
neutrino's journey from the core of the sun.
Kamiokande measures the high energy end of the $^8B$ neutrino spectrum, and
so assuming that the neutrino spectrum in $^8B$ beta decay in the center of
the sun is the same as that measured in the laboratory, one can determine
the number of $^8B$ neutrinos that the Homestake experiment should measure
with its lower energy threshold.
This calculation disagrees with the number measured at Homestake by at
least $2 \sigma$. It thus follows that the spectrum of $^8B$ neutrinos
detected on the earth differs from that in the core of the sun.

The Atmospheric Neutrino problem stems from a deficit in the ratio of
muon type to electron type neutrinos produced in the upper atmosphere
\cite{ATMEXPT1}.
When cosmic rays enter the upper atmosphere, they produce hadronic showers
containing many pions and kaons. Nearly all of the positively charged pions
decay via
\begin{equation}
\begin{array}{ll}
\pi^+ \rightarrow & \mu^+ + \nu_{\mu} \\
& ~\hookrightarrow e^+ + \nu_e + \bar\nu_{\mu}
\end{array}
\end{equation}
and the negatively charged pions decay via the charge conjugate to this
process. The majority of charged kaons either decay as the pions above
do, or they decay to pions which follow the above decay.

Notice that in the decay above, there are two $\nu_{\mu}$'s produced and
only one $\nu_e$ (where $\nu$ refers to both neutrinos and antineutrinos).
Naively we would expect that the ratio
$R_{calc} = \frac{\# \nu_\mu}{\# \nu_e}$ in the decay of many positively
and negatively charged pions and kaons would be equal to 2 \cite{GSB1}.
In more detailed treatments employing Monte Carlo programs to simulate
the hadronic showers in the upper atmosphere and the subsequent decay
processes (taking the effect of muon polarization into account), this ratio
is calculated to be about 1.8 \cite{ATMFLUX}. Measurements by
both IMB and Kamiokande find that the ratio is much lower,
and that the ratio of the measured to calculated ratios is
$R_{\mbox{\scriptsize{meas}}}/R_{\mbox{\scriptsize{calc}}} = 0.65$
with deviations of
about $0.1$ depending on the range of energy used
\cite{ATMEXPT1,ATMEXPT2}. However, there is still some question as to
whether or not there is such a deficit \cite{ATMEXPT3}.

There has been much activity in the past few years to explain both of these
problems (i.e. the Solar Neutrino and the Atmospheric Neutrino problems)
by the same minimal extension of
the standard model of particle physics. If we assume that neutrinos are
massive and that, like the quarks, their mass eigenstates differ from their
weak eigenstates, then the two sets of eigenstates are related by a
unitary matrix similar to the CKM matrix in the quark sector.
It is conventional to parameterize the oscillations between two
neutrinos by two parameters $\Delta m^2_0$ and $sin^2(2 \theta_0)$ where
$\Delta m^2_0 = m_{\nu_2}^2 - m_{\nu_1}^2$ is the difference in the squared
masses of the neutrinos and $\theta_0$ is the mixing angle.
Much work has been done in an attempt to account for the deficit of
solar neutrinos by oscillations of the electron neutrino with either
the muon or the tau neutrino as it travels through the sun.
These oscillations can be enhanced by the presence of matter.
The Atmospheric Neutrino problem can be explained by
oscillations between the muon neutrino and another neutrino
when it travels from the upper atmosphere to detectors several
km below the surface of the earth.
Experimental evidence puts quite strong constraints on what regions
of the parameter space are able to explain these experimental
measurements. However,
regions of $\Delta m^2_0$ and $sin^2(2 \theta_0)$ which account for the
Solar Neutrino problem do not overlap with those which account for
the Atmospheric Neutrino problem. It is then proposed that matter
enhanced $\nu_e - \nu_{\mu}$ oscillations account for the solar
neutrino problem, while $\nu_{\mu} - \nu_{\tau}$ vacuum oscillations
account for the deficit in the ratio
$R_{\mbox{\scriptsize{meas}}}/R_{\mbox{\scriptsize{calc}}}$
in atmospheric neutrinos.

One of the main difficulties with extracting information about
neutrino oscillations from the Solar and Atmospheric neutrinos
is that the initial spectrum and composition of the neutrinos
is not precisely known.
A laboratory based long baseline experiment has the advantage
of having a neutrino beam whose flux and spectrum can be measured
at the source (by putting a
small detector near the end of the muon decay tunnel).
By comparing the spectrum and composition
of the initial beam to that of the detected beam one can
unambiguously determine whether the beam has changed
while the neutrinos traveled from the source to the target. Also,
in an accelerator based experiment we have some control over the spectrum
and energy of the neutrino beam, and they can be tuned to investigate a
particular energy of interest.

The purpose of this paper is to study the theoretical
expectations for neutrino oscillations from a
long baseline accelerator experiment at KAON,
the proposed upgrade to TRIUMF. KAON will accelerate protons to
about 30 GeV, and will have a flux on the order of 140 $\mu$A providing
$8.74 \times 10^{14}$ protons on target per second. This high flux will
allow very long baseline neutrino oscillation experiments to be done.
Several such proposals are currently under active investigation by
the Neutrino Group at Triumf.
In this paper we begin by describing estimates for
the neutrino fluxes and spectra which could be expected
at KAON. We then compute numerically
the rate at which neutrinos would be measured a
large distance away for a variety of neutrino masses and
mixing angles and including the effect of matter (the MSW effect).
The  long baseline experiments which we have studied would either
confirm or rule out oscillations in
the region $\Delta m^2_0 = 10^{-3}$ to $10^1 eV^2$ and
$sin^2(2 \theta_0) > 0.4$. This is precisely the region which
is studied by the Atmospheric Neutrino measurements.
We also compute the minimum value of $\Delta m^2_0$
which can be explored with KAON for reasonable choice of the
baseline and the size of the detector and we estimate
 the values of $\Delta m^2_0$ and
$sin^2(2 \theta_0)$ which can be ruled out if no oscillations are found.

In section II, we briefly
review the theoretical framework for the study
of neutrino oscillations including both vacuum oscillations
and oscillations in matter (the MSW effect). In Section III we describe the
estimates of the neutrino fluxes expected for KAON and the methods used
to obtain these estimates. We begin Section IV by describing an
intermediate baseline experiment which has been proposed by the
Neutrino Group at TRIUMF \cite{neutgrp} using
KAON as a neutrino source and placing a detector 100 km away.
This baseline, which is larger than the 20 km baseline proposed
for the Brookhaven experiment is made possible by the very high neutrino
fluxes at KAON. To allow comparison
with the recent BNL-AGS proposal \cite{BNLAGS1,BNLAGS2}
we present results for  expected
event rates based on a 6300 tonne detector. We then proceed to our main result
namely the calculation of the event rates for
a very long baseline experiment in which the
neutrino beam from KAON would be sent through the earth to the SuperKamiokande
detector in Japan which is approximately 7200 km away. Matter effects are
important in this region for some range of the parameters.
Despite the low rates the extremely long baseline would allow much more
stringent limits on
$\Delta m^2_0$.

\section{Theoretical Background}

If neutrinos have a nonzero mass their mass eigenstates will not, in general,
coincide with the eigenstates that participate in weak interactions. These
two sets of eigenstates will are then
related by a unitary transformation. In the case of two neutrino flavors
with mass eigenstates $|\nu_1\rangle$ and
$|\nu_2\rangle$ and weak eigenstates $|\nu_\alpha\rangle$ and
$|\nu_{\beta}\rangle$ the eigenstates are related by a single mixing angle:

\begin{equation}
\left(
\begin{array}{c}
\nu_\alpha \\ \nu_\beta
\end{array}
\right) = \left(
\begin{array}{cc}
cos(\theta_0) & sin(\theta_0) \\
-sin(\theta_0) & cos(\theta_0)
\end{array}
\right) \left(
\begin{array}{c}
\nu_1 \\ \nu_2
\end{array}
\right)
\label{unitrans}
\end{equation}

The Schroedinger equation for propagation of a mass eigenstate of
momentum $k$ is given by:

\begin{equation}
i \frac{d}{dt} \left(
\begin{array}{c}
\nu_1 \\ \nu_2
\end{array}
\right) = \left(
\begin{array}{cc}
\sqrt{k^2+m_1^2} & 0 \\
0 & \sqrt{k^2+m_2^2}
\end{array}
\right) \left(
\begin{array}{c}
\nu_1 \\ \nu_2
\end{array}
\right) \approx
k \left(
\begin{array}{cc}
1 + \frac{m_1^2}{2 k^2} & 0 \\
0 & 1 + \frac{m_2^2}{2 k^2}
\end{array}
\right) \left(
\begin{array}{c}
\nu_1 \\ \nu_2
\end{array}
\right)
\end{equation}
where the last step is valid in the ultrarelativistic limit
$k^2 \gg m_i^2$. It is customary to subtract
$[k + (m_1^2 + m_2^2)/{4k}]$ times the identity from the Hamiltonian
(which has the effect of changing the overall phase of the wavefunction)
to obtain

\begin{equation}
i \frac{d}{dt} \left(
\begin{array}{c}
\nu_1 \\ \nu_2
\end{array}
\right) = \frac{1}{4 k} \left(
\begin{array}{cc}
- \Delta m_0^2 & 0 \\
0 & \Delta m^2_0
\end{array}
\right) \left(
\begin{array}{c}
\nu_1 \\ \nu_2
\end{array}
\right)
\end{equation}
where $\Delta m_0^2 = m^2_2 - m^2_1$. Using the transformation (\ref{unitrans})
this can be expressed in terms of the weak eigenstates as follows:

\begin{equation}
i \frac{d}{dt} \left(
\begin{array}{c}
\nu_\alpha \\ \nu_\beta
\end{array}
\right) = H \left(
\begin{array}{c}
\nu_\alpha \\ \nu_\beta
\end{array}
\right)
= \frac{\Delta m_0^2}{4 k} \left(
\begin{array}{cc}
- cos(2 \theta_0) & sin(2 \theta_0) \\
sin(2 \theta_0) & cos(2 \theta_0)
\end{array}
\right) \left(
\begin{array}{c}
\nu_\alpha \\ \nu_\beta
\end{array} \right)
\label{weakvac}
\end{equation}

It was first noted by Wolfenstein \cite{LW} that when electron neutrinos pass
through matter the $H_{ee}$ component of the Hamiltonian
 acquires an additional term. This term results from
the fact that  electron neutrinos can
scatter off the electrons in  matter via both  the charged and neutral
currents whereas
muon and tau neutrinos have only neutral current interactions with those
electrons. The term which is added to $H_{ee}$ is
$+ \sqrt{2} G_F N_e$, where $N_e$ is the electron density in the matter.
In the case of electron and muon neutrinos, and after subtracting
${\sqrt{2} G_F N_e}/{2}$ times the identity matrix from $H$,
equation (\ref{weakvac}) becomes:

\begin{equation}
\begin{array}{lll}
i \frac{d}{dt} \left(
\begin{array}{c}
\nu_e \\ \nu_\mu
\end{array}
\right) & = & {{\Delta m_M^2}\over{4 k} }\left(
\begin{array}{cc}
- cos(2 \theta_M) & sin(2 \theta_M) \\
sin(2 \theta_M) & cos(2 \theta_M)
\end{array}
\right) \left(
\begin{array}{c}
\nu_e \\ \nu_\mu
\end{array}
\right)
\end{array}
\label{weakmatter}
\end{equation}
where the mass difference $\Delta m_M^2$ and the  mixing angle $\theta_M$
in matter are
given by:

\begin{equation}
\Delta m^2_M = D_M \times \Delta m^2_0
\end{equation}
\begin{equation}
sin(2 \theta_M) = \frac{sin(2 \theta_0)}{D_M}
\end{equation}
and $D_M$ is expressed in terms of the electron number density $N_e$,
the vacuum oscillation length $L_0 = {4 \pi E \hbar c}/{\Delta m_0^2 c^4}$
and neutrino interaction length $L_e = {\sqrt{2} \pi \hbar c}/{G_F N_e}$
as follows:

\begin{equation}
D_M = \left[ 1 - 2 \left(\frac{L_0}{L_e}\right) cos(2 \theta_0) +
\left(\frac{L_0}{L_e}\right)^2 \right]^{\frac{1}{2}}
\end{equation}
The ratio ${L_0}/{L_e}$ in units relevant for this problem is given
by

\begin{equation}
\frac{L_0}{L_e} = 1.52 \times 10^{-4}
\left( \frac{E}{GeV} \right)
\left( \frac{eV^2}{\Delta m^2} \right)
\left(\frac{N_e}{2.1 mol/cm^3} \right)
\label{L0Le}
\end{equation}

Thus the effective mass difference and mixing angle are affected by the
presence of matter. This is the MSW effect.
If a $\nu_\mu$ is produced by some weak process, is
allowed to propagate (as a linear combination of the mass eigenstates), and
is then measured by some weak process a distance $L$ away, then the probability
that it will be measured as a $\nu_\mu$ is

\begin{equation}
P(\nu_\mu(0) \rightarrow \nu_\mu(L)) =
1 - sin^2(2 \theta_0) sin^2 \left( 1.27 \frac{\Delta m_0^2 L}{E} \right)  \\
\label{probnu}
\end{equation}
where $\Delta m^2$, $L$ and $E$ are measured
in $eV^2$, $km$ and $GeV$ respectively.

Note that the conventional application of the MSW effect is to
the Solar Neutrino problem in which electron neutrinos propagate
through a medium of variable density possibly passing through
a resonance at some point in their travels. The situation with
long baseline terrestrial experiments is significantly different.
In our case muon neutrinos travel through a region of
nearly constant density. The MSW effect will be appreciable only
for neutrinos whose {\bf energies} lie close to
the MSW resonant energy. This will occur only for a fraction of
the neutrinos in the beam.

\section{Neutrino Beam Calculations}

The neutrino group at TRIUMF \cite{neutgrp} has developed Monte Carlo methods
 for estimating
the neutrino fluxes at KAON using a Geant based program.
Neutrino fluxes and energy spectra are generated by a Monte Carlo simulation
of the magnetic horn focusing system used in E734 at Brookhaven \cite{Ahr86}.
The AGS beam at Brookhaven accelerates protons to 28.3 GeV, which is about
the same energy as would be expected at KAON. In the simulation, the
protons are incident upon a sapphire target rod where they produce
secondary hadrons.  The positively charged secondaries, which consist mostly of
pions and kaons, are focussed in the horns by fields of up to 5.3 T.  They then
enter a 200 m tunnel in which they decay to produce neutrinos. 3 km away the
neutrinos pass through the front face of a hypothetical detector.  Particle
tracking, decay and energy loss are simulated by the Monte Carlo routine GEANT
V3.15 \cite{Bru92}, while hadronic production is handled by GHEISHA
\cite{Car87} which is interfaced with GEANT.

Once a neutrino is produced by the decay of a hadron or muon it is assumed
that no interaction occurs before reaching the detector.  Therefore it is only
necessary to project the neutrino with its initial
momentum to the plane containing
the front face of the detector.  If the projected neutrino coordinates place it
inside an area of 8m by 8m at a distance of 3 km from the end of the last
focusing horn, the neutrino is accepted.   Momentum spectra for muon- and
electron-neutrinos and their anti-particles were collected based on a sample
of $1.2 \times 10^8$ protons incident upon the production target.
The neutrino energy spectra were then fitted to the smooth curves
shown in figure (\ref{KAONSPECT}). In the calculations which follow, we use
these fitted curves on the range from 1 to 6 GeV. At KAON, with a projected
proton current on target of 140 $\mu$A, the muon-neutrino flux at 3 km would
be $4.95 \times 10^5$ $\nu_{\mu}/cm^2/s$. Neutrino rates at
larger distances are estimated by assuming a $1/L^2$ dependence
of the neutrino flux. Monte Carlo flux calculations for detectors at
0.5, 1, 3, and 20 km indicate that the $1/L^2$ behavior is in
fact closely followed for distances beyond 1 km.

\section{Results}

There have recently been several proposals for long baseline neutrino
oscillation experiments using accelerator
\cite{BNLAGS1,BNLAGS2,JP,RHB-SJP,GF-BR,Nish}
and nuclear reactor \cite{RIS} neutrinos.
In a recent overview Parke \cite{SJP} has considered several possible long
baseline experiments. They fall into two main categories: experiments with
neutrinos on the order of 10's of GeV with baselines in the several 1000 km
range, and experiments with neutrinos on the order of several GeV with
baselines in the 10 to 100 km range. The higher energy neutrino beams allow
longer baselines because the neutrino scattering cross section is proportional
to its energy (which partly compensates for the ${1}/{L^2}$ reduction in
neutrino flux). However, as can be seen from equation (\ref{probnu}),
an increase in the neutrino energy requires
a proportionally longer baseline to probe the same $\Delta m^2_0$.
Because of its high flux, KAON offers a unique
opportunity to probe a very long (several thousand km) baseline with neutrino
energies peaked at about 1.5 GeV. Table \ref{accdetpair} compares
the expected event rates (for ordinary neutrino scattering events)
and the mass difference $(\Delta m^2)_{\lambda/2}$ for which the baseline is
exactly one half of an oscillation length at the peak energy of the
accelerator's neutrino beam. The table includes
two possible configurations using the KAON factory. In one experiment, which
has been proposed by the TRIUMF Neutrino Group, a
neutrino beam from KAON is directed towards a large detector (we use
6300 tonne as a reference)  which is
100 km from TRIUMF, and in the other the beam from KAON is directed towards
SuperKamiokande which is about 7200 km away. A crude estimate of
$\Delta m^2$ can be obtained by computing the distance to the first
minimum in the oscillation in equation (\ref{probnu}), so that
\begin{equation}
\left( \Delta m^2 \right)_{\lambda/2} = \frac{\pi}{2}
\left( \frac{E_{\nu,peak}}{1.27 L}
\right)
\end{equation}
In practice, if the neutrino flux is large, it is possible to
set more stringent limits on $\Delta m^2$ than those given above.
Note that although KAON and the AGS at Brookhaven have similar proton energies,
the peak energy of KAON is higher for a 200 meter decay
tunnel resulting in more high energy neutrinos.

\subsection{A 100 km Baseline}

It is clear from the results of this section that matter effects are negligible
for the region of parameter space which can be studied by a 100km baseline
experiment.
If neutrino oscillations are present part of the produced
$\nu_\mu$ beam will be converted either into $\nu_e$ or into
$\nu_\tau$. In this paper we shall consider only oscillations
between two neutrino species so that in our
calculations we assume that the $\nu_\mu$ converts $either$
into $\nu_e$ $or$ into $\nu_\tau$. (In the case of
$\nu_\mu \leftrightarrow \nu_\tau$ oscillations there
is never a matter effect.)

We now take the $\nu_\mu$ flux from the previous section
and use it to compute the rate of charged lepton production from
charged current exchange for various values of $\Delta m_0^2$ and
$sin^2(2\theta_0)$. The total cross sections for neutrino-nucleon
interactions summarized in Quigg \cite{Quigg} are
\begin{equation}
\frac{\sigma_T(\nu_l N \rightarrow l^- X)}{E}
= 6.75 \times 10^{-38}
\frac{cm^2}{GeV}
\end{equation}
\begin{equation}
\frac{\sigma_T(\bar{\nu_l} N \rightarrow l^+ X)}{E}
= 3.25 \times 10^{-38}
\frac{cm^2}{GeV}
\end{equation}
These inclusive
cross sections may be used if we are not interested in resolving the
neutrino energy but only in identifying the interacting neutrino.
(The usual method for determining the neutrino energy is to reconstruct
the events in the exclusive interaction $\nu_l p \rightarrow l^- n$. The
cross-section for this reaction flattens out at an energy of
about 1 Gev.)

Figures (\ref{KAON-BM1})(a) and (b) show contours of constant probability
that a neutrino or antineutrino produced at KAON would be measured as a
neutrino or antineutrino of the same flavor in a 6300 tonne detector at 100
km. (We chose a 6.3 kiloton detector to allow a more direct comparison
with the BNL proposal.)
Since there are no matter effects at this distance
the conversion rate for neutrinos is the same as the conversion rate for
antineutrinos and the contours in figures (\ref{KAON-BM1})(a) and (b) are
the same. In one year we expect about 490,000 events, so we may crudely
estimate the standard deviation in the number of events per year to be
$\sigma  = 100/\sqrt{490,000}\% \approx 0.14\%$. Then in one year we exclude
$\Delta m^2_0 \geq 1.3 \times 10^{-3}$ in the limit of large mixing and
$sin^2(2 \theta_0) \geq 0.01$ for $\Delta m^2_0 \geq 0.02 eV^2$ at the
$3 \sigma$ level. Figure
(\ref{KAON-BM1})(c) shows the rate of $\mu^-$ production in the 6300
metric tonne detector per year assuming a 65\% detector efficiency.
Figure (\ref{KAON-BM1})(d) shows the rate of
electron production assuming $\nu_\mu-\nu_e$ oscillations. Such oscillations
would result in electron appearance above the 3800 per year coming from
$\nu_e$ contamination in the beam if no oscillations were present. If the
$\nu_e$ beam could be well understood, then this appearance channel would
provide very concrete evidence of neutrino oscillations.
These results would improve on the present limits on $\Delta m^2_0$
by more than an order of magnitude, and cover the entire region allowed by
the Kamiokande and IMB atmospheric neutrino results.

Another important feature of the 100 km baseline is the
high event rate (as high as 1300 events per day using the total cross-section,
of which about 730 are $\nu_{\mu} n \rightarrow \mu^- p$).
The neutrino energy spectrum can be measured by reconstructing this latter
exclusive cross section, and the neutrino oscillation length is dependent
on the neutrino energy, as was previously
noted by Barger et al\cite{Bargeretal}. The exclusive cross section for
this process is approximately constant with
$\sigma(\nu_{\mu} n \rightarrow \mu^- p) = 9.0 \times 10^{-39} cm^2$
for energies on the range 1 to 6 GeV \cite{BNLAGS2}.
Thus if there $are$ neutrino oscillations, the dependence of the
neutrino deficit on energy may be used
as a more sensitive measure of the
values of $\Delta m^2_0$ and $sin^2(2 \theta_0)$.

In Figure (\ref{KAON-BM2}) we compare the expected rate of
neutrino events as a function of the neutrino energy with and without
oscillations. In \ref{KAON-BM2}(a), (c), and (e) we plot the probability
from equation (\ref{probnu}) as a function of energy
with $sin^2(2 \theta_0) = 0.25$ and $\Delta m_0^2 = 0.004$, $0.02$,
and $0.1$ $eV^2$ respectively.
Notice that there are minima in the event rates
whenever $E_n = 1.27 \Delta m^2 L \frac{2}{(2n+1) \pi}$ with
$n = 0, 1, 2, 3,...$. These minima correspond to ``coherent'' oscillations.
They occur at energies $E_0$, $\frac{E_0}{3}$,
$\frac{E_0}{5}$ and so on, corresponding to 1/2, 3/2, and 5/2
oscillation lengths. Note that only the
first few minima in the spectrum will be seen because as $n$ increases
the peaks at $E_n$ get closer together and the detector will
not be able to resolve them. In the low energy limit
the rate will be $1 - \frac{1}{2} sin^2(2 \theta_0)$.

In figures (\ref{KAON-BM2})(b), (d), and (f) we fold in the expected
KAON energy spectrum and then
compare the neutrino spectra which would be observed with and without
oscillations for the above values of $\Delta m_0^2$ and $sin^2(2\theta_0)$.
The rates shown are number of events per year assuming 65\% detector
efficiency and 1/2 GeV energy bins.

In summary, as has been previously noted \cite{neutgrp},
a 100 km baseline experiment at KAON can improve
the present limit on $\Delta m_0^2$ by more than an order of magnitude.
This is  sufficient to study {\it all} of the regime of interest
to the Atmospheric Neutrino experiments, and $\Delta m_0^2$ well below
the BNL-AGS proposal, but it is insufficient to say anything
about the Solar Neutrino experiments or matter effects.
The high event rates may allow a comparison of the energy spectrum of
the oscillating neutrinos.

\subsection{A 7200 km Baseline: KAON to SuperKamiokande}

A very exciting possibility would be to direct the KAON beam towards
the SuperKamiokande (SK) detector in Japan. The decay tunnel would have to
be aimed 35 degrees below the horizontal and the total length of the
baseline would be approximately 7200 km. KAON's high flux and relatively
low energy combined with SuperKamiokande's large size would allow a
probe of much lower values of
$\Delta m^2_0$ than other accelerator-detector
configurations. With no oscillations the rate for the KAON-SK experiment
would be
about 2 events per day assuming high efficiency in the 32,000 tonne
fiducial mass of SuperKamiokande (50,000 tonne total mass).

As in the case of the 100 km baseline experiment the idea is to do
a $\nu_\mu$ disappearance experiment.  The number of $\nu_\mu$
events at Kamiokande must be compared to the (presumably measured)
initial neutrino flux. The absence of
a reduction  in the number
of $\nu_\mu$ would set limits on $\Delta m_0^2$ and $sin^2(2 \theta_0)$.
If the oscillations are between $\nu_{\mu}$ and $\nu_{\tau}$  then
matter effects (the MSW effect) are not present.
Figures \ref{SKAM1}(a) and (b) show contours of constant probability for
neutrinos and antineutrinos $not$ oscillating (ie. being measured with
the same flavor with which they were produced) assuming vacuum oscillations.
Figures
\ref{SKAM1}(c) and (d) show the number of muons produced per year as a
function of $\Delta m_0^2$ and $sin^2(2 \theta_0)$. With approximately
700 events per year, an experiment which did not see a disappearance of
10\% of the muon neutrinos would rule out
$\Delta m_0^2 \geq 9 \times 10^{-5} eV^2$ in the limit of large mixing, and
$sin^2(2 \theta_0) \geq 0.2$ for any
$\Delta m^2_0 \geq 4 \times 10^{-4} eV^2$.

If the dominant oscillation is between $\nu_\mu$ and $\nu_e$ then
the calculation of the expected $\nu_\mu$ deficit must be
modified due to the matter effects. With this long baseline, the neutrinos
pass through a slowly changing density which varies with distance from
the center of the earth. In figure (\ref{SKAM2}) we show the
electron and muon neutrino components of the KAON neutrino beam at
SuperKamiokande calculated
assuming a constant density of $2.1$ $mol/{cm^3}$. These
results were found to be almost identical those those calculated by
numerically integrating equation (\ref{weakmatter})
with $\Delta m_M^2$ and $sin^2(2 \theta_M)$ varying according to the earth's
density profile. Figures \ref{SKAM2}(a) and (b) show the probability that
a neutrino and antineutrino arrive at SuperKamiokande with the same flavor
as they had when they were produced. Matter effects enhance the mixing of
neutrinos while they suppress the mixing of antineutrinos.

Figures \ref{SKAM2}(c) and (d) show the number of muons and electrons
which would be observed per year as a function of $\Delta m_0^2$ and
$sin^2(2 \theta_0)$. If no muon neutrino disappearance was observed at
a level of 10\%, this experiment could place limits of
$\Delta m_0^2 \geq 3 \times 10^{-4} eV^2$ for large mixing and
$sin^2(2 \theta_0) \geq 0.03$ if $\Delta m^2_0 \approx 10^{-3} eV^2$. This
latter mass difference has a resonance at the KAON peak energy of 1.5 GeV
when the neutrinos are passing through matter with electron density equal
to that of the earth. In addition to the muon disappearance, one could
look for electron appearance above the expected rate of about 5 per year
from $\nu_e$ contamination in the KAON beam. Although
this experiment is unable to probe the range of parameters
which are of interest in the Solar Neutrino problem, it would still be
very interesting because it probes a much lower neutrino mass difference
than any other experiment to date.

\begin{acknowledgments}
This work was supported in part by the Natural Sciences and Engineering
Council of Canada.
\end{acknowledgments}

%

%

%
\begin{table}
\caption{Parameters for various proposed long baseline neutrino
oscillation experiments. The Charged Current (CC) events per day
are given in the absence of neutrino oscillations. The parameter
$(\Delta m^2)_{\lambda/2}$ shown is the value of $\Delta m^2$ for
which a neutrino with energy equal to the peak energy of the
accelerator would just reach its first minimum at the detector.
Note that this $\Delta m^2$ is {\it not} necessarily the limit
which the experiment can probe. Other factors such as the event rate will
play an important role.}
\vspace{0.5cm}
\begin{tabular}{|l|c|c|c|c|c|}
\hline
Baseline & $E_{\nu,peak}$ & Baseline & Protons on Target
  & CC Events& $(\Delta m^2)_{\lambda/2}$ \\
& $(GeV)$ & $(km)$ & per second & per day & $(eV^2)$ \\
\hline
BNL-AGS & 1 & 20 & $3.5 \times 10^{13}$ & 320
& $6.2 \times 10^{-2}$ \\
KEK-SuperK \cite{note1} & 1 & 250 & $7.5 \times 10^{12}$ & 2.6
& $4.9 \times 10^{-3}$ \\
KAON 100 km & 1.5 & 100 & $8.74 \times 10^{14}$ & 1340
 & $1.8 \times 10^{-2}$ \\
FNAL-Soudan \cite{note2} & 10 & 710 & $2 \times 10^{13}$ & 9.5
& $1.7 \times 10^{-2}$ \\
CERN SPS-ICARUS & 6 & 730 & $8.3 \times 10^{12}$ & 29
& $1.0 \times 10^{-2}$ \\
KAON-SuperK & 1.5 & 7200 & $8.74 \times 10^{14}$ & 2
& $2.6 \times 10^{-4}$ \\
\hline
\end{tabular}
\label{accdetpair}
\label{tab2clg}
\end{table}

%
\begin{figure}
\caption{Monte Carlo calculation of neutrino and antineutrino spectra at KAON
using a 200 m decay tunnel. The spectra used in later calculations are the
fitted curves. Note that the raw fluxes generated by the Monte Carlo were
multiplied by 56.9 to convert them to $GeV^{-1} cm^{-2} s^{-1}$. Low
statistics then account for the noisy data in the electron antineutrino
spectrum.
\label{KAONSPECT}}
\end{figure}
\begin{figure}
\caption{Contours of constant probability of nonconversion of neutrino and
antineutrino flavors, and number of muons and electrons measured per year
by a 6300 tonne detector 100 km away assuming 65\% detector efficiency
as a function of neutrino oscillation parameters.
If no oscillations were present, then $N(\mu^-)$ would be
$4.9 \times 10^{5}$ and $N(e^-)$ would be 1340.
\label{KAON-BM1}}
\end{figure}
\begin{figure}
\caption{$P(\nu_\mu(0) \rightarrow \nu_\mu(L))$ and muon neutrino spectra
at 100 km assuming $sin^2(2 \theta_0) = 0.25$ and
$\Delta m^2_0 = 0.004$, $0.02$, and $0.1 eV^2$. If no oscillations
were present the probabilities in (a), (c), and (e) would be equal to unity.
The solid curves in equations (b), (d), and (f) are the spectra which would
be measured if no oscillations were present while the dashed curves are
the spectra if oscillations were present, and where the bin width is 1/2 GeV.
\label{KAON-BM2}}
\end{figure}
\begin{figure}
\caption{Contours of constant probability of nonconversion of neutrino and
antineutrino flavors and number of muons measured per year at SuperKamiokande
as a function of neutrino oscillation parameters assuming vacuum oscillations.
The rates assume high
efficiency in the 32,000 tonne fiducial mass. If no oscillations were
present, then $N(\mu^-)$ and $N(\mu^+)$ would be $700$ and $11$ respectively.
\label{SKAM1}}
\end{figure}
\begin{figure}
\caption{Contours of constant probability of nonconversion of neutrino
and antineutrino flavors and number of muons
and electrons measured per year at
SuperKamiokande as a function of neutrino oscillation parameters assuming
matter enhanced $\nu_e - \nu_\mu$ oscillations. The rates assume high
efficiency in the 32,000 tonne fiducial mass. If no oscillations were
present, then $N(\mu^-)$ and $N(e^-)$ would be $700$ and $5$
respectively.
\label{SKAM2}}
\end{figure}

\begin{references}
\bibitem{note1}
The numbers given are based on the upgrade to KEK discussed in \cite{Nish}.
\end{references}
\end{document}